\documentclass[twocolumn, times, twocolappendix]{aastex631}

\usepackage{amsmath}
\usepackage{txfonts}
\usepackage{xspace}
\usepackage{multirow}

\newcommand{\methanol}{CH$_3$OH\xspace}
\newcommand{\acetaldehyde}{CH$_3$CHO\xspace}
\newcommand{\methylformate}{CH$_3$OCHO\xspace}
\newcommand{\dimethylether}{CH$_3$OCH$_3$\xspace}

\shorttitle{Dimethyl Ether in MWC 480}
\shortauthors{Yamato et al.}
\graphicspath{{./}{figures/}}
\begin{document}

% \title{Dimethyl Ether Emission Tracing Icy Organic Sublimates in the MWC 480 Protoplanetary Disk}

% \title{Detection of Dimethyl Ether in the MWC 480 Protoplanetary Disk}

\title{Detection of Dimethyl Ether in the Central Region of the MWC~480 Protoplanetary Disk}

\author[0000-0003-4099-6941]{Yoshihide Yamato}
\affiliation{Department of Astronomy, Graduate School of Science, The University of Tokyo, 7-3-1 Hongo, Bunkyo-ku, Tokyo 113-0033, Japan}

\author[0000-0003-3283-6884]{Yuri Aikawa}
\affiliation{Department of Astronomy, Graduate School of Science, The University of Tokyo, 7-3-1 Hongo, Bunkyo-ku, Tokyo 113-0033, Japan}

\author[0000-0003-4784-3040]{Viviana V. Guzmán}
\affiliation{Instituto de Astrofísica, Pontificia Universidad Católica de Chile, Av. Vicuña Mackenna 4860, 7820436 Macul, Santiago, Chile}

\author[0000-0002-2026-8157]{Kenji Furuya}
\affiliation{Department of Astronomy, Graduate School of Science, The University of Tokyo, 7-3-1 Hongo, Bunkyo-ku, Tokyo 113-0033, Japan}
\affiliation{National Astronomical Observatory of Japan, Osawa 2-21-1, Mitaka, Tokyo 181-8588, Japan}

\author[0000-0003-2493-912X]{Shota Notsu}
\affiliation{Department of Earth and Planetary Science, Graduate School of Science, The University of Tokyo, 7-3-1 Hongo, Bunkyo-ku, Tokyo 113-0033, Japan}
\affiliation{Department of Astronomy, Graduate School of Science, The University of Tokyo, 7-3-1 Hongo, Bunkyo-ku, Tokyo 113-0033, Japan}
\affiliation{Star and Planet Formation Laboratory, RIKEN Cluster for Pioneering Research, 2-1 Hirosawa, Wako, Saitama 351-0198, Japan}

\author[0000-0002-2700-9676]{Gianni Cataldi}
\affiliation{National Astronomical Observatory of Japan, Osawa 2-21-1, Mitaka, Tokyo 181-8588, Japan}

\author[0000-0001-8798-1347]{Karin I.~\"Oberg}
\affiliation{Center for Astrophysics | Harvard \& Smithsonian, 60 Garden Street, Cambridge, MA 02138, USA}

\author[0000-0001-8642-1786]{Chunhua Qi}
\affiliation{Center for Astrophysics | Harvard \& Smithsonian, 60 Garden Street, Cambridge, MA 02138, USA}

\author[0000-0003-1413-1776]{Charles J.\ Law}
\altaffiliation{NASA Hubble Fellowship Program Sagan Fellow}
\affiliation{Department of Astronomy, University of Virginia, Charlottesville, VA 22904, USA}

\author[0000-0001-6947-6072]{Jane Huang}
\affiliation{Department of Astronomy, Columbia University, 538 W. 120th Street, Pupin Hall, New York, NY, United States of America}

\author[0000-0003-1534-5186]{Richard Teague}
\affiliation{Department of Earth, Atmospheric, and Planetary Sciences, Massachusetts Institute of Technology, Cambridge, MA 02139, USA}

\author[0000-0003-1837-3772]{Romane Le Gal}
\affiliation{Institut de Planétologie et d’Astrophysique de Grenoble (IPAG), Université Grenoble Alpes, CNRS, Grenoble, France}
\affiliation{Institut de Radioastronomie Millimétrique (IRAM), Saint-Martin d’Hères, France}

\correspondingauthor{Yoshihide Yamato}
\email{yyamato.as@gmail.com}



\begin{abstract}
% Complex organic molecules (COMs) in protoplanetary disks directly affect the chemical composition of forming planets and are thus key to understanding the chemical evolution from the interstellar medium (ISM) to planetary systems.
Characterizing the chemistry of complex organic molecules (COMs) at the epoch of planet formation provides insights into the chemical evolution of the interstellar medium (ISM) and the origin of organic materials in our Solar System.
% , yet observations of COMs in protoplanetary disks still remains scarce. 
% We present sensitive Atacama Large Millimeter/submillimeter Array observations of the disk around the Herbig Ae star MWC~480, where we detect dimethyl ether (\dimethylether) emission for the first time in this disk. 
We report a detection of dimethyl ether (\dimethylether) in the disk around the Herbig Ae star MWC~480 with the sensitive Atacama Large Millimeter/submillimeter Array observations. This is the first detection of \dimethylether in a non-transitional Class II disk.
The spatially unresolved, compact (${\lesssim}25$\,au in radius) nature, the broad line width (${\sim}30$\,km\,s$^{-1}$), and the high excitation temperature (${\sim}200$\,K) indicate sublimation of COMs in the warm inner disk. Despite the detection of \dimethylether, methanol (\methanol), the most abundant COM in the ISM, has not been detected, from which we constrain the column density ratio of \dimethylether/\methanol ${\gtrsim}7$. This high ratio may indicate the reprocessing of COMs during the disk phase, as well as the effect of the physical structure in the inner disk. We also find that this ratio is higher than in COM-rich transition disks recently discovered. This may indicate that, in the full disk of MWC~480, COMs have experienced substantial chemical reprocessing in the innermost region, while the COM emission in the transition disks predominantly traces the inherited ice sublimating at the dust cavity edge located at larger radii (${\gtrsim}20$\,au). 
% The parametric disk model analysis shows that this peculiar ratio can be explained by the different spatial distributions due to the different binding energies of \dimethylether and \methanol
% , where the beam dilution works differently
% . 
% We discuss the chemical origin of \dimethylether in the context of inheritance/reset of the protostellar chemistry and comparison with other Herbig disks, motivating the need for further observations towards a general picture of disk COM chemistry.  

\end{abstract}

%% \keywords{Classical Novae (251) --- Ultraviolet astronomy(1736) --- History of astronomy(1868) --- Interdisciplinary astronomy(804)}


\section{Introduction} \label{sec:intro}
Complex organic molecules (COMs), defined empirically as carbon-bearing molecules with six or more atoms, serve as precursors of prebiotic molecules and thus provide insight into the origin of organic materials and life in our solar system \citep[e.g.,][]{Ceccarelli2023}. COMs are formed initially on the cold grain surfaces in dark clouds \citep[e.g.,][]{Herbst2009} and will be readily observable in radio wavelengths in the warm (${\gtrsim}100$\,K) environment, where they are released into the gas phase by thermal desorption. Indeed, observations with the Atacama Large Millimeter/submillimeter Array (ALMA) have detected a number of COMs in the warm inner envelopes around protostars \citep[e.g.,][]{Jorgensen2018, Belloche2020, Yang2021} and the warm protostellar disk around the FU Ori type outbursting star V883 Ori \citep{vantHoff2018, Lee2019, Yamato2024}. COMs have also been detected in solar system comets, for example, 67P/Churyumov-Gerasimenko, which has been extensively studied by the Rosetta mission \citep{Altwegg2019}. While comparative studies of COM abundance ratios suggest that COMs in comets are at least partially inherited from the interstellar medium \citep[ISM;][]{Drozdovskaya2019}, characterizing the distributions and abundances of COMs in protoplanetary disks is crucial for understanding the chemical evolution from the ISM to planetary systems.

% However, it still remains a challenge to observe sublimated COMs in old ($>1$\,Myr) protoplanetary disks due to their colder nature. COMs are hidden in ices on dust grain surfaces in the cold midplane (${\sim}20$\,K) and their sublimation regions are limited only within inner a few au in the disks around typical T Tauri stars. The weak, extended emission of the most abundant COM, methanol (\methanol), has been detected in the outer cold region (${\gtrsim}30$\,au) of the nearest planet-forming disk TW~Hya, but its low abundance relaive to H$_2$ (${\sim}10^{-11}$--$10^{-12}$) suggest the inefficient non-thermal desorption origin \citep{Walsh2016}. Complex nitrile CH$_3$CN are also detected in a handful of disks \citep[e.g.,][]{Oberg2015}, but the low rotational temperatures (${\sim}30$--50\,K) based on the excitation analysis of multiple $k$-ladder transitions indicate the non-thermal origin and/or gas-phase formation \citep{Loomis2018, Bergner2018, Ilee2021}.

Recent sensitive observations with ALMA have revealed the presence of thermally desorbed oxygen-bearing COMs in a few warm transition disks around Herbig Ae stars. \citet{vanderMarel2021} detected warm methanol (\methanol) gas localized at the asymmetric dust trap in the transition disk around IRS~48, followed by additional detections in the transition disks of HD 100546 and HD 169142 \citep{Booth2021_CH3OH, Booth2023_HD169142}. IRS~48 exhibits the highest chemical complexity among these disks, where dimethyl ether (\dimethylether), methyl formate (\methylformate), and ethylene oxide ($c$-H$_2$COCH$_2$) have been detected in addition to \methanol \citep{Brunken2022, Booth2024_IRS48}. Other oxygen-bearing molecules such as SO, SO$_2$ and NO are also detected at the dust trap, indicating the presence of oxygen-rich gas and a low C/O ratio of $<1$ \citep{Booth2021, Leemker2023}. HD~100546 also exhibits \methylformate emission but no \dimethylether and $c$-H$_2$COCH$_2$, possibly suggesting a different reservoir of COMs compared to IRS 48 \citep{Booth2024_HD100546}. In these transition disks around Herbig Ae stars, the high luminosity of the central star and the inner dust cavity provide a warmer disk condition, which makes it possible for COMs to sublimate.

However, the current limited detections of sublimated COMs in transition disks preclude a comparison of the COM reservoir between other Herbig and T Tauri disks. Although weak, extended emission of \methanol has been detected in the outer cold region (${\gtrsim}30$\,au) of the nearest T Tauri disk TW~Hya, its low abundance relative to H$_2$ (${\sim}10^{-11}$--$10^{-12}$) suggests a non-thermal desorption origin \citep{Walsh2016}. Similar attempts to detect \methanol in the disks around Herbig Ae stars HD~163296 and MWC~480 yielded non-detections (HD~163296; \citealt{Carney2019}) or tentative detections (MWC~480; \citealt{Loomis2018}) in the outer cold region. In this paper, we present the first detection of \dimethylether in the MWC~480 protoplanetary disk. MWC~480 is a ${\sim}7$\,Myr old \citep{Montesinos2009} Class~II Herbig Ae star with a stellar mass of $M_\star\approx2.1\,M_\odot$ \citep{Simon2019, Teague2021_MAPS} in the Taurus-Auriga star-forming region ($d\approx162$\,pc; \citealt{GaiaColaboration2018}).
% , a bolometric luminosity of $L_\star \approx 21.9\,L_\odot$ \citep{Montesinos2009}, and a mass accretion rate of $\dot{M} \approx 1.3 \times10^{-7}\,M_\odot$\,yr$^{-1}$ \citep{Mendigutia2013}. 
The MWC 480 disk, which has been observed in the Molecules with ALMA at Planet-forming Scales (MAPS) ALMA Large Program \citep{Oberg2021}, is one of the most extensively studied disks around Herbig Ae stars. The MWC~480 disk exhibits a prominent dust ring at ${\approx}98$\,au \citep{Long2018, Liu2019, Sierra2021}, and the velocity perturbation observed in CO and its isotopologue emission suggests the presence of a planet at ${\approx}245$\,au \citep{Teague2021_MAPS, Izquierdo2023}. The outer disk (${\gtrsim}30$\,au) of MWC~480 is characterized by a gas-phase C/O $>1$ \citep{Bosman2021_MAPSVII, Jiang2023} and abundant hydrocarbons and cyanides \citep{Ilee2021, Guzman2021}, while the inner disk chemistry is unknown.

We describe the ALMA Band 7 observations in Section \ref{sec:observation}. We present the detection of \dimethylether, the tentative detection of \methylformate, and the non-detection of \methanol, from which we constrain the excitation condition and column densities based on the spectral analysis in Section \ref{subsec:CH3OCH3_detection}. We discuss the results in Section \ref{sec:discussion}, and finally summarize the study in Section \ref{sec:summary}.

\section{Observations} \label{sec:observation}
%%%%%%%%%%%%%%%%% Observtaional details of 2021.1.00535.S %%%%%%%%%%%%%%%%%%%%%%%%
We observed the MWC~480 disk in Band 7 during ALMA Cycle 8 (project code: 2021.1.00535.S; PI: Y.\ Yamato). Observations were carried out in five execution blocks. Observational details including observation date, number of antennas, on-source integration time, mean precipitable water vapor (PWV), baseline coverage, angular resolution, maximum recoverable scale (MRS), and calibrator information are summarized in Table \ref{tab:observations}. 

% The correlator setups included ten spectral windows (SPWs) with Frequency Division Mode (FDM). One of the SPWs covered a wide bandwidth of 1875\,MHz and had a spectral resolution of 1.1\,MHz ($\approx$\,1.1\,km s$^{-1}$), which was originally dedicated for continuum acquisition and was used in this study. The other SPWs were used to target other molecular lines, which will be presented in future works.

%%%%%%%%%%%%%%%%% calibration %%%%%%%%%%%%%%%%%%%%%%%%
The pipeline calibrations were performed by ALMA staff using the standard ALMA pipeline. Subsequent self-calibration and imaging were performed using the Common Astronomy Software Applications \citep[CASA;][]{CASA} version 6.2.1.7. The continuum visibilities were produced by averaging the line-free channels specified by visually inspecting the data cubes delivered from the observatory. The continuum emission peak on the image from each execution block was aligned to a common direction, $\alpha(\mathrm{ICRS}) = 04^{\mathrm{h}}58^{\mathrm{m}}46\fs279, \delta(\mathrm{ICRS}) = +29^{\mathrm{d}}50^{\mathrm{m}}36\fs40$, which was used as the disk center in the following analysis. Then, five rounds of phase-only self-calibration and one round of amplitude self-calibration were performed, resulting in a signal-to-noise ratio (S/N) increase by a factor of ${\approx}15$. The solutions were then applied to the entire dataset. The continuum emission was finally subtracted from the line visibilities by fitting a first-order polynomial to line-free channels using the CASA task \texttt{uvcontsub}.

%%%%%%%%%%%%%%%%% imaging %%%%%%%%%%%%%%%%%%%%%%%%
We focused on the spectral window originally dedicated for continuum acquisition, which covered a frequency range of 298.88--300.75\,GHz and had a spectral resolution of 1.1\,MHz (${\approx}1.1$\,km s$^{-1}$). This spectral window contained a number of \dimethylether and \methylformate transitions. We generated a single image cube for the entire frequency range by imaging the self-calibrated, continuum-subtracted visibilities with a channel spacing of 4\,km\,s$^{-1}$ and a Briggs robust value of 0.5 in the modified Briggs weighting scheme of \texttt{briggsbwtaper} implemented in the \texttt{tclean} task. The large channel width of 4 km s$^{-1}$ intended to boost the S/N of detected \dimethylether emission while resolving its broad line width. The CLEANed image cube has a synthesized beam of $0\farcs31\times0\farcs21$ (PA $=8\fdg2$) and a root mean square (RMS) noise level of 0.33 mJy beam$^{-1}$ at 4\,km\,s$^{-1}$ channel width. The RMS noise was measured on the line-free region of the image cube prior to primary beam correction. We also generated the continuum image at 1.03\,mm with a Briggs robust value of $-0.5$ to better resolve the gap-ring structure, which is shown in the top left panel of Figure \ref{fig:observation_summary}. Additionally, we report the detection of a new submillimeter source in the same field of view in Appendix \ref{appendix:new_submm_detection}.

%%%%%%%%%%%%%%%%% Observtaional details of 2021.1.00982.S %%%%%%%%%%%%%%%%%%%%%%%%
We also used the Band 7 data from another project (project code: 2021.1.00982.S; PI: V.\ V.\ Guzman). Observations were carried out in three execution blocks and their details are summarized in Table \ref{tab:observations}. The continuum spectral window of this data set, which is used in this study, covered a frequency range of 288.56--290.43\,GHz with a spectral resolution of 1.1\,MHz (${\approx}1.2$\,km s$^{-1}$). This spectral window covered a cluster of \methanol transitions, which was used to constrain the upper limit on the \methanol column density. We followed the same self-calibration procedure as above, but with CASA version 6.4.1. We performed three rounds of phase-only self-calibration and one round of amplitude self-calibration, which resulted in an increase in S/N by a factor of ${\approx}5.5$. We generated the image cube of the continuum spectral window in the same manner as above. The resulting beam size and the RMS noise level were $0\farcs31\times0\farcs22$ (P.A. $= -3\fdg4$) and 0.67 mJy beam$^{-1}$ at 4 km\,s$^{-1}$ channel width.

%%%%%%%%%%%%%%%% Absolute flux calibration unceratnity %%%%%%%%%%%%%%%
For both datasets, the absolute flux calibration uncertainty is expected to be ${\lesssim}10$\% based on the Quality Assurance (QA2) criteria described in ALMA Cycle 8 Technical Handbook. The following analyses consider only the statistical uncertainty and does not include the systematic flux calibration uncertainty.  

\begin{deluxetable*}{lcccccccc}
\label{tab:observations}
% \tablewidth{\textwidth}
\tablecaption{Observational Details}
\tablehead{\colhead{Date} & \colhead{\# of Ant.} & \colhead{On-source Int.} & \colhead{PWV}  & \colhead{Baseline} & \colhead{Ang. Res.} & \colhead{MRS} & \multicolumn{2}{c}{Calibrators}\\
\cline{8-9}
\colhead{} & \colhead{} & \colhead{(min)} & \colhead{(mm)} & \colhead{(m)} & \colhead{($\arcsec$)} & \colhead{($\arcsec$)} & \colhead{Bandpass/Amplitude} & \colhead{Phase}}
\startdata
\multicolumn{9}{c}{2021.1.00535.S (PI: Y. Yamato)} \\
\hline
2022 Aug. 8 & 42 & 44 & 0.7 & 15--1302 & 0.2 & 2.6 & J0510+1800 & J0438+3004 \\
2022 Aug. 9 & 40 & 44 & 0.2 & 15--1246 & 0.2 & 3.0 & J0510+1800 & J0438+3004 \\
2022 Aug. 9 & 40 & 44 & 0.1 & 15--1261 & 0.2 & 3.1 & J0510+1800 & J0438+3004 \\
2022 Aug. 10 & 42 & 44 & 0.4 & 15--1302 & 0.2 & 2.8 & J0510+1800 & J0438+3004 \\
2022 Aug. 10 & 41 & 44 & 0.5 & 15--1302 & 0.2 & 3.1 & J0510+1800 & J0438+3004 \\
\hline
\multicolumn{9}{c}{2021.1.00982.S (PI: V. V. Guzman)} \\
\hline
2022 Aug. 11 & 45 & 40 & 0.5 & 15--1302 & 0.2 & 2.9 & J0510+1800 & J0438+3004 \\
2022 Aug. 11 & 44 & 40 & 0.5 & 15--1302 & 0.2 & 2.8 & J0510+1800 & J0438+3004 \\
2022 Aug. 14 & 47 & 40 & 0.6 & 15--1302 & 0.2 & 2.8 & J0510+1800 & J0438+3004 \\
\enddata
\end{deluxetable*}

\section{Detection of Dimethyl Ether}\label{subsec:CH3OCH3_detection}
% \subsection{Matched filtering}
\subsection{Spectra and Spatial Distributions}
We first extracted the integrated spectra by spatially integrating the image cubes in a circular aperture with 0\farcs6 diameter toward the disk center, which securely covers the apparent emission as shown later. 
% The uncertainty of the spectrum is estimated by an empirical way in the extracted spectrum: first we calculated the standard deviation of the spectrum using all channels, and then iteratively estimated the uncertainty by excluding the channels with values larger than two times the standard deviation at each step. 
The middle panel of Figure \ref{fig:observation_summary} shows the resulting spectrum of the data containing detected \dimethylether transitions. We found a clear spectral feature at ${\sim}299.89$\,GHz, whose line center spectrally coincides with the rest frequencies of \dimethylether transitions taking into account the spectral shift due to the systemic velocity (5.1\,km,s$^{-1}$; \citealt{Pietu2007, Teague2021_MAPS}). This feature is a blend of 12 individual \dimethylether transitions (see Table \ref{tab:spectroscopic_data} in Appendix \ref{appendix:LTE_model} for their spectroscopic properties) and shows a peak S/N of 9.4. In addition, two tentative features are seen at ${\sim}299.12$\,GHz and ${\sim}300.06$\,GHz, which also match with the frequencies of \dimethylether and \methylformate transitions. These features show peak S/Ns of 3.4 and 4.7, respectively. Because they are lower than typical detection criteria of $5\sigma$ and the latter is blended between different species (\dimethylether and \methylformate), these transitions were considered tentative detections. The zoomed-in spectra around these features are presented in the bottom panels of Figure \ref{fig:observation_summary} along with the expected spectral location of the \dimethylether and \methylformate transitions. We note that an exhaustive search for other molecular lines that could be responsible for those features were performed, but only \dimethylether and \methylformate can consistently explain these features among the lines of detectable species/transitions in typical disk environment. 
% These features coincide with the expected locations of multiple transitions, and are found to be blends of them. 
% We also extracted the spectra around these features by matched filtering method \citep{Loomis2018_MF}, which is shown in Figure \ref{fig:spectrum_fit_mf} (bottom row). This method extract the response spectra by applying a filter of a certain spatio-kinematic pattern on the visibility plane (see \citealt{Loomis2018_MF} for technical details). Since we do not know the exact spatial/velocity distribution of the emission a priori, we used a simple Keplerian-rotating disk filter with an outer radius of 0\farcs15. While the feature at $\sim$\,299.89\,GHz shows a filter response of $\sim$\,16$\sigma$, other two features represent $\sim$\,4--5$\sigma$ responses, further confirming the (tentative) detection of these features.

The velocity-integrated intensity maps of these spectral features, generated using the Python package \texttt{bettermoments} \citep{bettermoments} with no flux threshold for pixel inclusion, are shown in the top panels of Figure \ref{fig:observation_summary} along with the 1.03~mm dust continuum image. These maps show the spatially-unresolved, compact emission with a peak S/N of $\sim$\,5--12. 

Several strong transitions of \methanol and \acetaldehyde were covered by the data from the project 2021.1.00982.S (see Table \ref{tab:spectroscopic_data} in Appendix \ref{appendix:LTE_model} for covered strong transitions), but they have not been detected in the image plane.
% Figure \ref{} in Appendix \ref{} presents the spectrum covering a number of \methanol transitions, exhibiting their non-detections. 

We also verified the (tentative) detections/non-detections of these COMs by matched filter analysis \citep{Loomis2018_MF}. The details and resulting response spectra are shown in Appendix \ref{appendix:MF}. The ${\sim}5$--$17\sigma$ responses for \dimethylether and \methylformate signify the detection of them, although it should be noted that the detected features are the blend of multiple transitions.

% As filter kernel, we used a simple Keplerian mask generated with the Python script \texttt{keplerian\_mask} \citep{Teague2020_kepmask} assuming the disk inclination angle of $37\fdg0$, position angle of $148\fdg0$ and central stellar mass of $2.11\,M_\odot$ \citep{Oberg2021, Teague2021_MAPS}. The outer radius of the mask was set as $0\farcs15$ based on the extent of the emission in the velocity-integrated intensity maps. The Python package \texttt{VISIBLE} \citep{Loomis2018_MF} was used to compute the cross-correlation between observed visibilities and the filter kernel. The resultant response spectra are shown in Figure \ref{fig:MF_response} in Appendix \ref{appendix:MF}. The brightest cluster of \dimethylether lines shows a ${\sim}17\sigma$ filter response, while the other two features exhibit ${\sim}5$--6$\sigma$ filter responses. 

% We also show the integrated spectra of the data containing bright \methanol transitions in Figure \ref{fig} in Appendix \ref{}, which exhibit the nondetection of \methanol. These \methanol transitions are also observed in other COM-rich disks and detected \citep[e.g.,][]{Booth2021_CH3OH, Booth2024_HD100546, Booth2024_IRS48}. No detections are also found in the matched filter analysis (Figure \ref{fig:MF_response}).

% These emissions are all spatially unresolved with $\sim0\farcs3$ beam and confined to the inner region ($\lesssim$\,20\,au radius).

\subsection{Spectral Analysis}\label{subsec:spectral_analysis}
To characterize the excitation temperatures and column densities of molecules, we employ a spectral fit with a simple local thermodynamic equilibrium (LTE) slab model. We basically followed the method described in \citet{Yamato2024}, which is detailed in Appendix \ref{appendix:LTE_model}. 
% Briefly, considering four molecular species (\dimethylether, \methylformate, \methanol and \acetaldehyde), we fit the entire spectra simultaneously. Although \methanol and \acetaldehyde are not detected, this simultaneous fit allows us to constrain the upper limit of their column densities. 
Briefly, the entire spectra are modeled with a total of seven free parameters: beam-averaged column densities of four molecular species ($N$(\dimethylether), $N$(\methylformate), $N$(\methanol) and $N$(\acetaldehyde)), excitation temperature $T_\mathrm{ex}$, which is assumed to be common among different species/transitions, line width $\Delta V$ (including the broadening due to disk rotation), and the systemic velocity $v_\mathrm{sys}$. Although \methanol and \acetaldehyde are not detected, this simultaneous fit allows us to constrain the upper limit on their column densities. The assumption of a common excitation temperature for all COM species reflects the expectation that different COMs sublimate simultaneously with water ice. 

We explored the parameter space with the affine-invariant Markov Chain Monte Carlo (MCMC) algorithm implemented in the Python package \texttt{emcee} \citep{emcee}. The details of this fitting procedure are described in Appendix \ref{appendix:LTE_model}. Figure \ref{fig:observation_summary} compares the best-fit model with the observed spectra. The three observed spectral features are well reproduced by the best-fit model. While the robust feature at $\sim$\,299.89\,GHz is explained by the cluster of \dimethylether transitions with relatively low upper state energies of 36--45\,K, other two tentative features correspond to the transitions of \dimethylether and \methylformate with higher upper state energies of 187--447\,K (see Table \ref{tab:spectroscopic_data}). 
% The full list of the detected transitions, including their spectroscopic properties, can be found in Appendix \ref{}. 
The result of the fit is summarized in Table \ref{tab:fit_result}. The excitation temperature and line width are constrained to be $T_\mathrm{ex} = 200_{-50}^{+50}$\,K and $\Delta V = 30_{-6}^{+8}$\,km\,s$^{-1}$, where the uncertainties are 16th and 84th percentiles (corresponding to $1\sigma$) of the posterior distributions. 
% The high excitation temperature and broad line width indicate that the emission originates from the inner disk where COMs have sublimated and the Keplerian velocity is larger. 
The resulting beam-averaged column densities of \dimethylether and \methylformate are $7.4_{-3.2}^{+4.6}\,\times\,10^{15}$ and $7.4_{-4.9}^{+3.6}\, \times\,10^{14}$ cm$^{-2}$, respectively. The upper limits on the column densities of \methanol and \acetaldehyde are also constrained to be $<9.1\,\times\,10^{14}$ and $<1.9\,\times\,10^{14}$ cm$^{-2}$, respectively. These upper limits are 99.85th percentile (corresponding to $3\sigma$) of the posterior distributions. From these constraints on column densities, we derived a \dimethylether/\methanol column density ratio (lower limit) of ${>}7$.
% , which is significantly higher than the typical values in protostellar envelopes ($\sim10^{-2}$-- less than unity; e.g., \citealt{Yang2021}). 
% We also constrain the \methylformate/\methanol ratio to be $>0.0074$. Although the best-fit value of \methylformate column density and the upper limit of \methanol column density suggests the ratio is larger than unity, the long tail in the posterior probability distribution of the \methylformate column density due to the low S/N results in a weak constraint.  

\begin{figure*}
% \epsscale{1.15}
\plotone{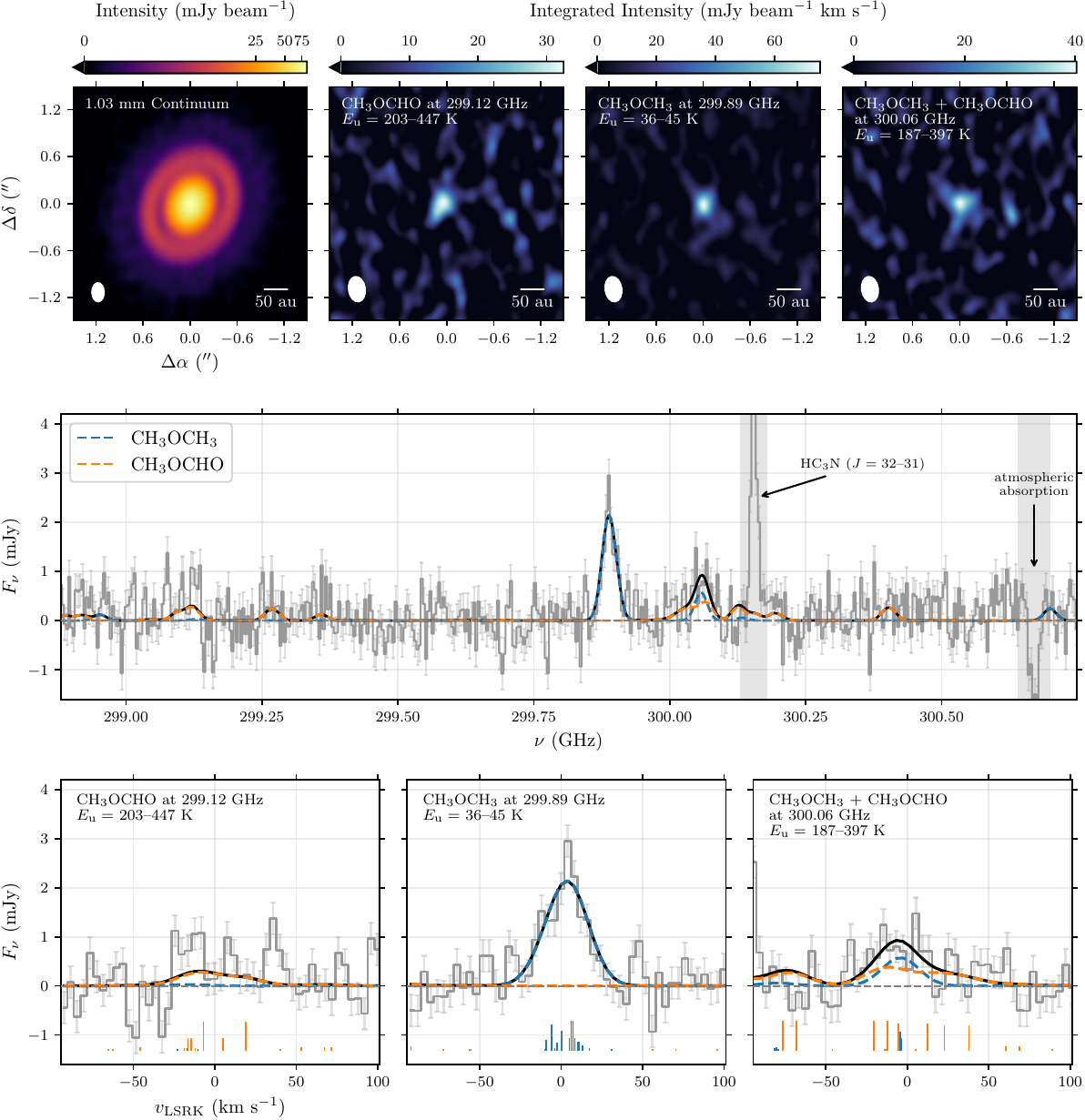}
\caption{Summary of the observations toward the MWC~480 disk. Top: 1.03 mm continuum image (left) and velocity-integrated intensity maps of (tentatively) detected COMs, CH$_3$OCH$_3$ and CH$_3$OCHO (others). The peak S/N of the right three maps are 5.4, 12, and 6.0, respectively. The white ellipse in the lower left corner indicates the synthesized beam. A scale bar of 50 au is shown in the lower right corner. Middle: Spectra extracted in a circular aperture with 0\farcs6 diameter toward the disk center (gray), overlaid with the best-fit spectral model for \dimethylether (blue), \methylformate (orange), and sum of both (black). The semitransparent error bars represent the associated 1$\sigma$ uncertainties. The frequency ranges removed from the fit due to the atmospheric absorption or contamination from other molecular lines are shaded by gray. Bottom: Zoom-in view of the middle panel in the vicinity of the (tentatively) detected features but with the spectral axis in velocity with respect to a reference frequency of 299.11170\,GHz (left), 299.89209\,GHz (middle), and 300.05338\,GHz (right), respectively. The vertical line segments in the bottom of the panels indicate the center and the relative strength (at 200\,K) of each transition. 
% The detected features are heavily blended with multiple transitions, and the channel width do not separate them due to the spectrally crowded nature and the wide line widths of these transitions.
}
\label{fig:observation_summary}
\end{figure*}

\begin{deluxetable}{lccccc}
\label{tab:fit_result}
\tablecaption{Result of the Spectral Analysis}
\tablehead{\colhead{Molecule} & \colhead{$T_\mathrm{ex}$} & \colhead{$\Delta V_\mathrm{FWHM}$} & \colhead{$v_\mathrm{sys}$}& \colhead{$N$} & \colhead{$/N(\mathrm{CH_3OH})$} \\
\colhead{} & \colhead{(K)} & \colhead{(km s$^{-1}$)} & \colhead{(km s$^{-1}$)} & \colhead{(cm$^{-2}$)} & \colhead{}}
\startdata
\dimethylether & \multirow{4}{*}{$200_{-50}^{+50}$} & \multirow{4}{*}{$30_{-6}^{+8}$} & \multirow{4}{*}{$4.3^{+2.4}_{-2.5}$} & $7.4_{-3.2}^{+4.6} \times 10^{15}$ & $> 7$ \\
\methylformate & & & & $7.4_{-4.9}^{+3.6} \times 10^{14}$ & $> 0.81^\dagger$\\
\methanol & & & & $< 9.1\times10^{14}$ & --- \\
\acetaldehyde & & & & $< 1.9\times10^{14}$ & --- \\
\enddata
\tablenotetext{\dagger}{The median value of $N$(\methylformate) divided by $N$(\methanol) upper limit, which should be viewed as tentative. See Appendix \ref{appendix:LTE_model} for details.}
\end{deluxetable}

% \subsection{}

% \begin{itemize}
%     \item mom0 and spectra
%     \item spectrum fitting with XCLASS?
%     \item estimates on column density and (if possible) temperature
%     \item Or we can use the temperature based on MAPS reference model
%     \item CH3OCH3/CH3OH ratio (lower limit)
% \end{itemize}

\section{Discussion} \label{sec:discussion}
\subsection{Chemical Origin of the COM Emission}\label{subsec:chemical_origin_of_COMs}
% Here, we discuss the chemical origin of the observed COM emission based on our analyses. 
The spatially unresolved, compact (${\lesssim}25$\,au radius) \dimethylether emission with a broad line width (${\sim}30$\,km\,s$^{-1}$) and a high excitation temperature (${\sim}200$\,K; see Section \ref{subsec:spectral_analysis}) indicates warm \dimethylether gas in the inner disk. This is the first observational evidence of COMs sublimated in the inner warm region of the MWC~480 disk. The observed broad line width is also reproduced with the emission originating from ${\lesssim}20$--30\,au radii, where the disk temperature is ${\gtrsim}80$--100\,K, by a more detailed radiative transfer calculation (see Appendix \ref{appendix:toy_model}), further supporting this interpretation. Indeed, the peak brightness temperature of CO $J=2$--1 emission already reaches ${\gtrsim}70$\,K at ${\approx}50$\,au \citep{Law2021}, implying that the inner disk of MWC~480 is warm enough (${\gtrsim}100$\,K) for COMs to sublimate. Furthermore, super-resolution image of the Band 6 continuum emission shows a steep brightness increase inside 20\,au radius \citep{Yamaguchi2024}, potentially suggesting the sublimation of water and COMs there that modify the dust optical properties. As the (sub-)mm dust continuum emission from the midplane is optically thick in the inner ${\lesssim}40$\,au \citep{Sierra2021}, the observed \dimethylether emission likely traces the disk surface layer.

% The wide line width is also reproduced with the more realistic disk model (Section \ref{sec:toy_model}), further supporting this interpretation.   
% This unusually wide line width is reproduced with the emission from the innermost region ($\lesssim10$\,au) in the toy model analysis (Section \ref{sec:toy_model}). The dust temperature in the model reaches $\sim80$--$100$\,K in the disk surface ($z/r\sim0.2$--0.3) at $\sim10$\,au (see Figure \ref{fig:model_comparison}; \citealt{Zhang2021}). These temperatures well agree with the sublimation temperature of \dimethylether calculated from the binding energies ($\approx4000\pm500$\,K; ref}) at gas densities of $\sim10^8$--$10^{10}$ cm$^{-3}$ (see e.g., \citealt{Furuya2014}), suggesting that the observed emission traces the thermally desorbed ice inside the snow line. 
% The tentatively-detected \methylformate can also be fitted with the same excitation temperature, implying that the observed \methylformate emission also traces the thermally desorbed component, although this should be confirmed by the future observations with higher sensitivities.

The \dimethylether detection and its icy origin lead us to expect the detection of other COMs, particularly the most abundant COM in the ISM, \methanol. However, the \methanol transitions ($E_\mathrm{u}{\gtrsim}$40\,K) have not been detected in the present observations,
% \footnote{\citet{Loomis2020} tentatively detected \methanol lines at a coarser angular resolution ($\sim1\arcsec$), but this likely traces the cold component in the outer disk given the low upper state energies of the lines (5--20\,K).}
resulting in a high \dimethylether/\methanol column density ratio of ${>}7$ (Section \ref{subsec:spectral_analysis}). 
% Given that the observed emission is not spatially resolved, one of the possible interpretations for this peculiar ratio is the difference in their emitting region sizes; \methanol emission may originate from more inner region and beam-diluted more severely. 
This value is considerably higher than the typical values in the warm inner envelopes around protostars \citep[e.g.,][]{Jorgensen2018, Yang2021} by more than one order of magnitude. This may indicate the efficient formation of \dimethylether during the disk phase. While it is anticipated that \methanol predominantly forms via hydrogenation of CO on the cold grain surfaces in the dark cloud \citep[e.g.,][]{Watanabe2002}, more complex species, such as \dimethylether and \methylformate, may also form on the grain surface in-situ in disks. Disk chemical models suggest that radical-radical reactions on the warm (${\sim}30$--50\,K) grain surfaces may enhance the abundance of \dimethylether and \methylformate \citep[e.g.,][]{Furuya2014};
\begin{align}
    \mathrm{CH_3 + CH_3O} &\longrightarrow \mathrm{CH_3OCH_3}, \\
    \mathrm{HCO + CH_3O} &\longrightarrow \mathrm{CH_3OCHO},
\end{align}
where the ingredient radicals (CH$_3$, CH$_3$O, and HCO) are the photodissociation products of \methanol ice \citep{Oberg2009}. These reactions, triggered by energetic processes such as UV, X-rays, and cosmic-rays, will play a pivotal role in determining the abundance of COMs during the disk phase. Additionally, gas-phase formation in the inner warm region may also be a contributing factor, as suggested by astrochemical models \citep{Taquet2016, Garrod2022}.

The derived column density ratio (\dimethylether/\methanol${>}7$) is still higher than the prediction of those models, which is lower than unity \citep[e.g.,][]{Furuya2014, Taquet2016}. This implies that some additional effect is necessary to explain the observations.
% Therefore, the larger column density of \dimethylether than that of \methanol implies some additional effects. 
One possibility is the difference in the distributions between \dimethylether and \methanol.
% due to the difference in sublimation temperature (or binding energy). 
The slightly higher binding energy of \methanol (${\approx}$5000--6600\,K; \citealt{Ferrero2020, Minissale2022}) than that of \dimethylether (${\approx}$4000--6500\,K; \citealt{Ligterink2023}) may cause the \methanol ice to sublimate in warmer regions (i.e., smaller radii) and the \methanol emission to be more heavily beam-diluted (see Appendix \ref{appendix:toy_model} for a test of this possibility). 
% Considering this difference in the emitting (or sublimation) regions, we demonstrated that at least a moderate \dimethylether/\methanol abundance ratio of ${\gtrsim}0.3$ is necessary to explain the observations by a radiative transfer calculation of a disk model (Appendix \ref{}). 
% We demonstrate that the observed high column density ratio can be explained by an 
Additionally, dust continuum emission may selectively hide the \methanol emission that could originate from a more inner region
%, if the dust-emitting disk midplane is warmer than the line-emitting disk surface at the \methanol emitting radii due, for example, to viscous accretion heating 
\citep[e.g.,][]{Bosman2021, Yamato2024}.

Taken together, we suggest that the substantial reprocessing of COMs during the disk phase, in conjunction with an additional effect of the difference in spatial distributions could explain the detection of \dimethylether but not \methanol. Further observations at higher resolution and sensitivity are required to more accurately constrain the distributions and abundance ratio of COMs, thus enabling a better inference of the reprocessing of COMs during the disk phase. 

\subsection{Comparison to Other Disks}\label{subsec:comparison_to_other_disks}
\subsubsection{COM-rich Transition Disks}
\begin{figure}
\epsscale{1.15}
\plotone{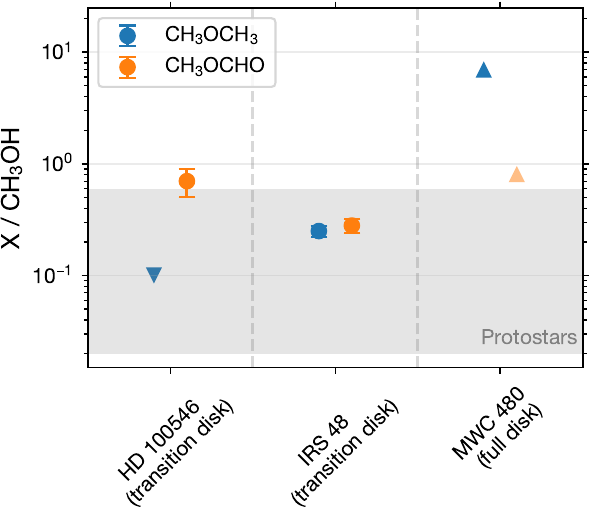}
\caption{Comparison of the column density ratios of \dimethylether and \methylformate relative to \methanol in three disks, HD~100546 \citep{Booth2024_HD100546}, Oph~IRS~48 \citep{Booth2024_IRS48}, and MWC 480 (this work). The upward and downward triangles indicate lower and upper limits, respectively. Given the tentative detection (see Section \ref{subsec:spectral_analysis}), the value of \methylformate in MWC~480 is plotted in transparent. 
% The \dimethylether value in MWC~480 based on the disk modeling considering different emitting regions (Appendix \ref{}) is also plotted in transparent. 
The shaded region indicates the typical range of protostellar values measured by extensive line surveys (PILS; \citealt{Jorgensen2016}, CALYPSO; \citealt{Belloche2020}, and PEACHES; \citealt{Yang2021}).}
\label{fig:abundance_comparison}
\end{figure}

Recent deep line surveys have identified the emission from the warm gas of sublimated COMs in three transition disks: HD~100546, Oph~IRS~48, and HD~169142 \citep{Booth2023_HD169142, Booth2024_HD100546, Booth2024_IRS48}. Multiple species have been detected in Oph~IRS~48 (\methanol, \dimethylether, \methylformate, and $c$-H$_2$COCH$_2$) and HD~100546 (\methanol and \methylformate), while only \methanol has been detected in HD~169142. On the other hand, in full disks, no previous detections of sublimated COMs have been reported, and the present study provides the first evidence of icy organic sublimates. Figure \ref{fig:abundance_comparison} compares the abundance ratios of \dimethylether and \methylformate relative to \methanol among HD~100546 \citep{Booth2024_HD100546}, Oph~IRS~48 \citep{Booth2024_IRS48}, and MWC~480. The abundance ratios exhibit a considerable degree of variability among the sources, with particularly elevated values in the full disk of MWC~480. This suggests the potentially different chemistry and/or the effect of varying physical structures between transition and full disks. The abundance ratios in these transition disks may be determined predominantly by the sublimation of inherited ices at the dust cavity edge (or dust trap) located at larger radii \citep[${\gtrsim}20$\,au; e.g.,][]{Booth2021_CH3OH}. In contrast, in the full disk of MWC~480, COM emission is likely to originate from more inner regions (${\lesssim}20$--30\,au), where COMs may have experienced more reprocessing by e.g., X-ray-driven chemistry and/or gas-phase chemistry in the disk surface layer. Indeed, the higher fractional abundance of HCO$^+$ relative to CO in MWC~480 compared to HD~100546 and IRS~48 may indicate the higher X-ray flux \citep{Aikawa2021, Booth2024_HD100546, Booth2024_IRS48}, although the lower HCO$^+$ abundance in HD~100546 and IRS~48 could be attributed to the presence of gas-phase H$_2$O that destroys HCO$^+$ \citep[e.g.,][]{Leemker2021}.
% Indeed, there is an observational evidence that the stellar X-ray is 
% Additionally, the emission of COMs in Oph~IRS~48 is co-located with the asymmetric dust crescent, 

It should also be noted that we obtain a tight upper limit on \acetaldehyde column density, which is less than ${\sim}3$\% of \dimethylether column. This is in a similar trend to the non-detection of \acetaldehyde in the IRS~48 and HD~100546 disks \citep{Booth2024_IRS48, Booth2024_HD100546}, while in contrast to the protostellar envelopes/disks where \acetaldehyde exists in a comparable amount of \dimethylether \citep[e.g.,][]{Jorgensen2018, Yang2021, Yamato2024}. This might indicate that the specific reactions and/or branching ratios which are relevant to \acetaldehyde formation/destruction may be at work in the disk phase, as suggested by the detection of $c-$H$_2$COCH$_2$, the isomer of \acetaldehyde, in IRS~48 \citep{Booth2024_IRS48}.

\subsubsection{Non-detection in HD~163296}

It is also noteworthy that the same transitions of \dimethylether toward another well-studied full disk around the Herbig Ae star HD~163296 have not been detected with a similar sensitivity and resolution in the same project\footnote{It should be noted that the physical scale traced by these observations are slightly different between MWC~480 and HD~163296 since the angular resolutions (${\sim}0\farcs3$) are the same but the distances are different (162 pc vs. 101 pc).}. This suggests that the chemistry and/or physical structure in the inner disk may be different between MWC~480 and HD~163296.
% while they are similar in the outer disk as observed, for example, in the MAPS Large Program \citep{Zhang2021, Ilee2021, Guzman2021}.
The CO snowline in MWC~480 is located at a larger radius (${\sim}100$\,au) than in HD~163296 (${\sim}65$\,au) as demonstrated in a comprehensive thermochemical modeling study \citep{Zhang2021}, implying a warmer inner disk of MWC~480. The higher rotation temperature of the HC$_3$N emission in the inner ${\lesssim}10$\,au (${\sim}90$\,K vs. ${\sim}60$\,K) may also indicate that the inner region is warmer in MWC~480 \citep{Ilee2021}. Furthermore, the slightly higher mass accretion rate and bolometric luminosity of MWC~480 \citep{Fairlamb2015, Montesinos2009, Mendigutia2013} may result in a greater quantity of organics being delivered to the inner disk \citep[e.g.,][]{Banzatti2023}.

\section{Summary and Conclusion} \label{sec:summary}
We present ALMA Band 7 observations toward the disk around the Herbig Ae star MWC~480. We detected the line emission of \dimethylether and tentatively detected \methylformate for the first time in the MWC~480 disk. This is the first detection of \dimethylether in a non-transitional Class II disk. The compact, spatially unresolved nature at a resolution of ${\sim}0\farcs3$ (or ${\sim}25$\,au radius), the broad line width (${\sim}30$ km\,s$^{-1}$ in FWHM), and the high excitation temperature (${\sim}200$\,K) indicate ice sublimation in the warm inner disk. 
% We also conducted the parametric disk modeling to confirm that the \dimethylether emission from the $\lesssim25$\,au region well reproduces the observed wide line width with a more realistic disk structure. 

Furthermore, we present a non-detection of \methanol, which results in a high \dimethylether/\methanol column density ratio of ${>}7$. This high ratio may be explained by the efficient formation of \dimethylether during the disk phase triggered by UV, X-ray, and cosmic rays, with a possible difference in the spatial distributions between \dimethylether and \methanol. We compare the column density ratios of COMs with those in COM-rich transition disks that have recently been discovered. The higher ratios in the full disk of MWC~480 may point to substantial chemical reprocessing of COMs in the innermost region, while those in transition disks may predominantly trace the sublimation of inherited ice at the dust trap located at ${\gtrsim}20$\,au.

% despite the \dimethylether detection. The disk model analysis showed that the varying sublimation temperature (and thus emitting radius) between \methanol and \dimethylether can explain this peculiar non-detection. The constrained \dimethylether/\methanol ratio of $\gtrsim0.07$ is comparable to the ice abundance ratios theoretical disk chemical models predicted and to the values in other evolutionary stages, suggesting the inheritance of the ice from the early phases. Alternatively, it may be also possible that the in-situ formation in the disk is responsible for the observed COMs. 

The detection of \dimethylether in MWC~480 provides the first evidence of icy organic sublimates in a full disk. Further observations, including the confirmation of the detection with other transitions, and detailed disk chemical modeling are necessary to better understand the inner disk chemistry. In particular, future JWST observations will provide complementary information about the COM abundances in gas and solid phases, which can be compared with the gas-phase abundances estimated from ALMA observations.
% , from which we can do a comparative discussion about the gas-ice chemistry in the inner disk.

% \begin{acknowledgments}
% \textcolor{red}{We thank...}

% \begin{acknowledgments}
\section*{Acknowledgements}
We thank the anonymous referee for constructive comments that improved the manuscript. This paper makes use of the following ALMA data: ADS/JAO.ALMA\#2021.1.00535.S and ADS/JAO.ALMA\#2021.1.00982.S. ALMA is a partnership of ESO (representing its member states), NSF (USA) and NINS (Japan), together with NRC (Canada), NSTC and ASIAA (Taiwan), and KASI (Republic of Korea), in cooperation with the Republic of Chile. The Joint ALMA Observatory is operated by ESO, AUI/NRAO and NAOJ. The National Radio Astronomy Observatory is a facility of the National Science Foundation operated under cooperative agreement by Associated Universities, Inc. 
Y.Y. acknowledges support by Grant-in-Aid for the Japan Society for the Promotion of Science (JSPS) Fellows (KAKENHI Grant Number JP23KJ0636).
Y.A. acknowledges support by MEXT/JSPS Grants-in-Aid for Scientific Research (KAKENHI) Grant Numbers JP20H05847, JP21H04495, and JP24K00674 and NAOJ ALMA Scientific Research Grant code 2019-13B.
V.V.G gratefully acknowledges support from FONDECYT Regular 1221352, and ANID CATA-BASAL project FB210003.
K.F. acknowledges support by MEXT/JSPS Grants-in-Aid for Scientific Research (KAKENHI) Grant Number JP21K13967.
S.N.~is grateful for support from Grants-in-Aid for JSPS (Japan Society for the Promotion of Science) Fellows Grant Number JP23KJ0329, and MEXT/JSPS Grants-in-Aid for Scientific Research (KAKENHI) Grant Numbers JP20K22376, JP23K13155, JP20H05845, and JP23H05441.
Support for C.J.L. was provided by NASA through the NASA Hubble Fellowship grant No. HST-HF2-51535.001-A awarded by the Space Telescope Science Institute, which is operated by the Association of Universities for Research in Astronomy, Inc., for NASA, under contract NAS5-26555.
% \end{acknowledgments}

% \end{acknowledgments}


\vspace{5mm}
\facilities{ALMA}


\software{CASA \citep{CASA}, bettermoments \citep{bettermoments}, VISIBLE \citep{Loomis2018_MF}, \texttt{keplerian\_mask.py} \citep{Teague2020_kepmask}, emcee \citep{emcee}, \texttt{corner.py} \citep{corner}, RADMC-3D \citep{RADMC3D}, Numpy \citep{Numpy}, Scipy \citep{Scipy}, Matplotlib \citep{Matplotlib}, Astropy \citep{astropy:2013, astropy:2018, astropy:2022}, cmocean \citep{cmocean}}



\appendix

\section{Matched Filter Analysis}\label{appendix:MF}
\begin{figure*}
% \epsscale{1.15}
\plotone{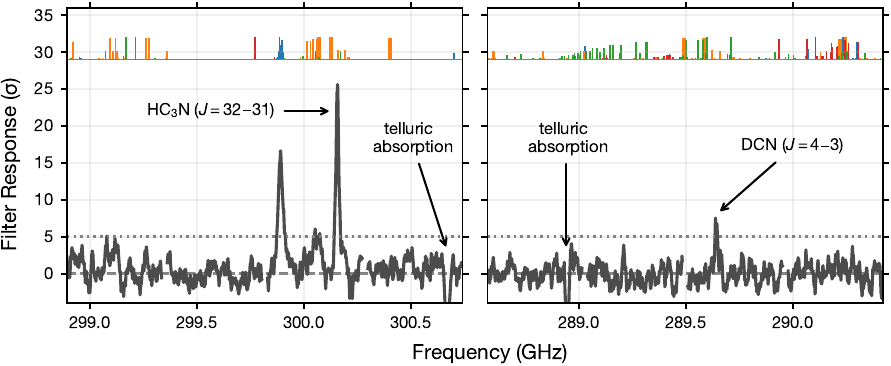}
\caption{Matched filter responses. The horizontal dotted line marks $5\sigma$. The relative intensities of covered lines at 200\,K are shown by vertical segments for \dimethylether (blue), \methylformate (orange), \methanol (red), and \acetaldehyde (green). A cluster of \dimethylether lines at ${\sim}299.89$\,GHz exhibits a ${\sim}17\sigma$ response, while tentative features at ${\sim}299.12$\,GHz and ${\sim}300.06$\,GHz shows ${\sim}5$--$6\sigma$ responses.}
\label{fig:MF_response}
\end{figure*}

We performed the matched filter analysis \citep{Loomis2018_MF}, which is commonly used to extract weak molecular line emission in protoplanetary disks, to verify the detections/non-detections of COMs. As a filter kernel, we used a simple Keplerian-rotating disk model generated with the Python script \texttt{keplerian\_mask} \citep{Teague2020_kepmask} assuming a disk inclination angle of $37\fdg0$, position angle of $148\fdg0$ and central stellar mass of $2.1\,M_\odot$ \citep{Oberg2021, Teague2021_MAPS}. The outer radius of the disk was set as $0\farcs15$ based on the extent of the emission in the velocity-integrated intensity maps. The Python package \texttt{VISIBLE} \citep{Loomis2018_MF} was used to compute the cross-correlation between observed visibilities and the filter kernel. The resultant response spectra are shown in Figure \ref{fig:MF_response}. The brightest cluster of \dimethylether lines at ${\sim}299.89$\,GHz shows a ${\sim}17\sigma$ filter response, while the two tentative features at ${\sim}299.12$\,GHz and ${\sim}300.06$ GHz exhibit ${\sim}5$--6$\sigma$ filter responses. The response spectra do not exhibit any ${\ge}5\sigma$ responses of \methanol and \acetaldehyde transitions.

\section{Details of Spectral Analysis}\label{appendix:LTE_model}
Here, we describe the details of the spectral analysis (Section \ref{subsec:spectral_analysis}) to derive the excitation temperature and column density of COMs. We fit the entire spectra by an LTE slab model based on the basic radiative transfer equations \citep[e.g.,][]{Yamato2024}. The emergent intensity of the line emission $I_\nu$ at each frequency is modeled as 
\begin{equation}
    I_\nu = (B_\nu(T_\mathrm{ex}) - B_\nu(T_\mathrm{CMB}))(1 - e^{-\tau_\nu}),
\end{equation}
where $B_\nu$ is the Planck function of the blackbody radiation, $T_\mathrm{ex}$ is the excitation temperature, $T_\mathrm{CMB} = 2.73$\,K is the temperature of cosmic microwave background, and $\tau_\nu$ is the optcial depth profile. We assume that $T_\mathrm{ex}$ is shared by different species of COMs, i.e., the emitting regions are common among different COMs, given that COMs are expected to sublimate at a similar temperature. The optical depth profile $\tau_\nu$ is computed as 
\begin{equation}\label{eq:tau}
    \tau_\nu = \sum_i \frac{c^2A_{\mathrm{ul},i}N_{\mathrm{u},i}}{8\pi\nu^2}\left[\exp\left(\frac{h\nu}{k_\mathrm{B}T_\mathrm{ex}}\right) -1\right]\phi_{\nu, i},
\end{equation}
where $c$ is the speed of light, $A_{\mathrm{ul}, i}$ is the Einstein A coefficient of $i$th transition, $N_{\mathrm{u}, i}$ is the upper state column density of $i$th transition, $h$ is the Planck constant, $k_\mathrm{B}$ is the Boltzmann constant, and $\phi_{\nu, i}$ is the line profile function of $i$th transition. The summation $i$ goes over different transitions of all molecular species. The upper state column density $N_{\mathrm{u}, i}$ is related to the total column density $N$ as
\begin{equation}
    N_{\mathrm{u}, i} = \frac{N}{Q(T_\mathrm{ex})}g_{\mathrm{u}, i}\exp\left(-\frac{E_{\mathrm{u}, i}}{k_\mathrm{B}T_\mathrm{ex}}\right),
\end{equation}
where $Q$ is the partition function, $g_\mathrm{u}$ is the upper state degeneracy, and $E_\mathrm{u}$ is the upper state energy. The line profile function $\phi_{\nu, i}$ in Equation \ref{eq:tau} is defined as 
\begin{equation}\label{eq:lpf}
    \phi_{\nu, i} = \frac{1}{\sqrt{2\pi}\sigma_{\nu, i}}\exp\left[-\frac{(\nu - \nu_{0, i} - \delta\nu_i)^2}{2\sigma_{\nu, i}^2}\right],
\end{equation}
where $\sigma_{\nu, i}$ is the frequency line width, $\nu_{0, i}$ is the rest frequency of $i$th transition, and $\delta\nu_i$ is the frequency shift due to the systemic velocity $v_\mathrm{sys}$. The frequency line width $\sigma_{\nu, i}$ is related to the velocity line width in Full Width Half Maximum (FWHM) $\Delta V_\mathrm{FWHM}$ as $\sigma_{\nu, i} = \dfrac{\nu_{0, i}}{c}\dfrac{\Delta V_\mathrm{FWHM}}{\sqrt{8\ln 2}}$\footnote{This sort of column density retrieval, including the functions of querying molecular spectroscopic data, is packaged into a public Python code \texttt{specfit} (\url{https://github.com/yyamato-as/specfit}).}. 

Within this model, we consider $>600$ transitions of four molecular species, \dimethylether, \methylformate, \methanol and \acetaldehyde, whose Einstein A coefficient is larger than $10^{-8}$\,s$^{-1}$ and upper state energy is lower than $1000$\,K. Table \ref{tab:spectroscopic_data} lists the strong transitions and their spectroscopic properties that effectively constrain the column density (and its upper limit). The spectroscopic data of these transitions, including partition functions, are taken from the Cologne Database for Molecular Spectroscopy \citep[CDMS;][]{CDMS1, CDMS2, CDMS3} for \dimethylether and \methanol and the Jet Propulsion Laboratory (JPL) database \citep{JPL} for \methylformate and \acetaldehyde. In total, we include seven free parameters into the fit: the excitation temperature $T_\mathrm{ex}$, the velocity line width $\Delta V_\mathrm{FWHM}$, the systemic velocity $v_\mathrm{sys}$, and the logarithm of column densities of these four species. We note that here we assume a Gaussian line shape (Equation \ref{eq:lpf}) and the line width is the one after including the broadening due to Keplerian rotation rather than the intrinsic one. This predominantly reflects the underlying assumption that the emission is optically thin, where only the velocity-integrated intensity matters for column density derivation. The actual line shape may be different from Gaussian as usually seen in the spectra of rotating disks, but given the unresolved emission, the low S/N, and the low spectral resolution, we do not perform a detailed line shape modeling here and it is partly addressed in the radiative transfer modeling (Appendix \ref{appendix:toy_model}). 

\begin{figure*}
\epsscale{1.2}
\plotone{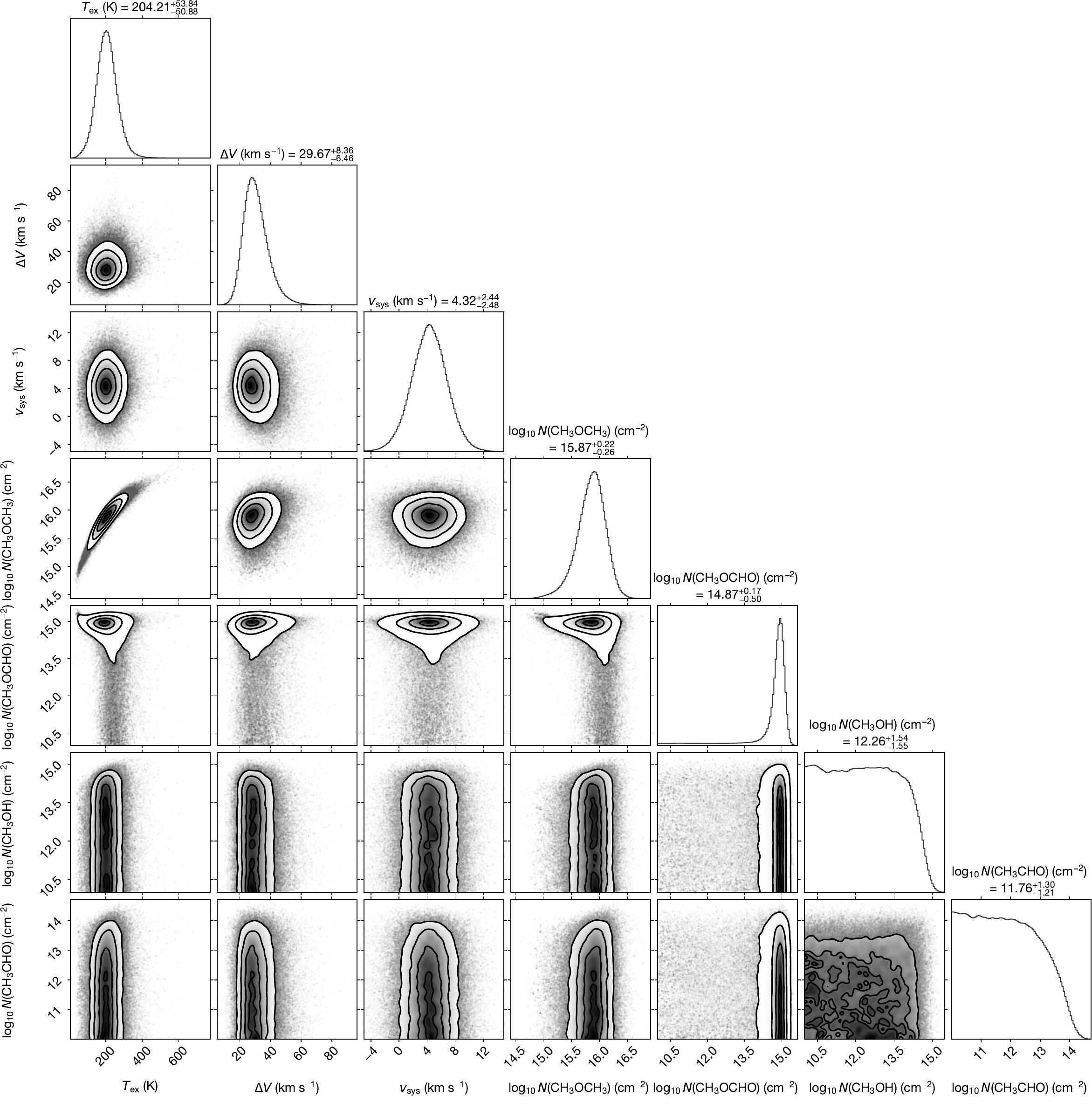}
\caption{Parameter covariances and marginalized posterior distributions for the MCMC spectral fit. The values and errors in the top of each panel indicate 16th, 50th (median), and 84th percentiles of posterior distributions.}
\label{fig:corner}
\end{figure*}

We explore the parameter space with the affine-invariant Markov Chain Monte Carlo (MCMC) algorithm implemented in the Python package \texttt{emcee} \citep{emcee} using uniform prior distributions:
\begin{align*}
    T_\mathrm{ex} \, \mathrm{(K)} &= \mathcal{U}(5, 1000),\\
    \Delta V_\mathrm{FWHM}\,\mathrm{(km\,s^{-1})} &= \mathcal{U}(0.0, 100.0), \\
    v_\mathrm{sys}\,\mathrm{(km\,s^{-1})} &= \mathcal{U}(-5.0, 15.0), \\
    \log_{10}N(\mathrm{X}) \, \mathrm{(cm^{-2})} &= \mathcal{U}(10.0, 20.0), 
\end{align*}
where $\mathcal{U}(a, b)$ denotes the uniform distributions between $a$ and $b$, and X represents the four molecular species. We exclude the following frequency ranges from the fit due to the contamination from other molecular lines and telluric absorption lines: [288.85, 289.00], [289.62, 289.66], [300.13, 300.18], and [300.64, 300.70] in GHz. We run 50 walkers for 10000 steps to sample the posterior distributions and the initial 2000 steps are discarded as burn-in. Figure \ref{fig:corner} shows the parameter covariances and marginalized posterior distributions. Only the upper limits on \methanol and \acetaldehyde column density are constrained due to the non-detection. The median value of the systemic velocity (${\approx}4.3$ km s$^{-1}$) is slightly shifted from the literature value (${\approx}5.1$ km s$^{-1}$; \citealt{Pietu2007, Teague2021_MAPS}) due to the coarse channel width (${\approx}4$ km s$^{-1}$), and these are consistent within the uncertainty. From these posterior samples, we compute the column density ratios relative to \methanol, where $N$(\dimethylether)/$N$(\methanol) is constrained to be ${>}7$ (0.15th percentile, corresponding to $3\sigma$ for a Gaussian distribution). For \methylformate, while the 0.15th percentile of $N$(\dimethylether)/$N$(\methanol) is very small due to the long tail in the posterior distributions of $N$(\methylformate) (see Figure \ref{fig:corner}), the median value of $N$(\methylformate) divided by the upper limit of $N$(\methanol) are listed in Table \ref{tab:fit_result} and added in Figure \ref{fig:abundance_comparison} as a reference. For \acetaldehyde, we do not obtain meaningful constrains due to the non-detection. 

% We also verified that the emission is optically thin. The optical depth profile calculated from the median values of posterior distributions are all less than 0.003 as shown in Figure \ref{fig:optical_depth}. 

\startlongtable
\begin{deluxetable*}{lccccc}
\label{tab:spectroscopic_data}
\tablecaption{Spectroscopic Properties of Major Molecular Lines}
\tablehead{\colhead{Transition} & \colhead{$\nu_0$ (GHz)} & \colhead{$\log_{10} A_\mathrm{ul}$ (s$^{-1}$)} & \colhead{$g_\mathrm{u}$} & \colhead{$E_\mathrm{u}$ (K)}}
\startdata
\multicolumn{5}{c}{Dimethyl Ether (\dimethylether)} \\
\hline
5$_{4,2}$--4$_{3,1}$ AE & 299.884130 & $-3.763$ & 22 & 36.1 \\
5$_{4,2}$--4$_{3,2}$ EA & 299.886128 & $-3.769$ & 44 & 36.2 \\
5$_{4,1}$--4$_{3,1}$ EA & 299.886804 & $-3.769$ & 44 & 36.1 \\
5$_{4,1}$--4$_{3,2}$ AE & 299.888802 & $-3.763$ & 66 & 36.1 \\
5$_{4,1}$--4$_{3,1}$ EE & 299.889910 & $-3.784$ & 176 & 36.1 \\
5$_{4,2}$--4$_{3,2}$ EE & 299.890981 & $-3.784$ & 176 & 36.1 \\
5$_{4,2}$--4$_{3,1}$ AA & 299.892089 & $-3.763$ & 66 & 36.1 \\
5$_{4,1}$--4$_{3,2}$ AA & 299.896762 & $-3.763$ & 110 & 36.1 \\
8$_{3,5}$--7$_{2,6}$ AE & 299.899470 & $-4.007$ & 34 & 45.4 \\
8$_{3,5}$--7$_{2,6}$ EA & 299.899781 & $-4.008$ & 68 & 45.4 \\
8$_{3,5}$--7$_{2,6}$ EE & 299.903025 & $-4.008$ & 272 & 45.4 \\
8$_{3,5}$--7$_{2,6}$ AA & 299.906424 & $-4.007$ & 102 & 45.4 \\
29$_{6,23}$--29$_{5,24}$ AA & 300.062299 & $-3.825$ & 590 & 447.5 \\
29$_{6,23}$--29$_{5,24}$ EE & 300.062707 & $-3.825$ & 944 & 447.5 \\
29$_{6,23}$--29$_{5,24}$ AE & 300.063115 & $-3.825$ & 354 & 447.5 \\
29$_{6,23}$--29$_{5,24}$ EA & 300.063116 & $-3.825$ & 236 & 447.5 \\
\hline
\multicolumn{5}{c}{Methyl Formate (\methylformate)} \\
\hline
23$_{5,18}$--22$_{5,17}$ E, $v_t = 0$ & 299.097870 & $-3.410$ & 94 & 183.9 \\
25$_{3,22}$--24$_{3,21}$ E, $v_t = 1$ & 299.111695 & $-3.406$ & 102 & 387.6 \\
23$_{5,18}$--22$_{5,17}$ A, $v_t = 0$ & 299.123858 & $-3.410$ & 94 & 183.9 \\
27$_{1,26}$--26$_{2,25}$ A, $v_t = 1$ & 299.128806 & $-4.243$ & 110 & 397.2 \\
27$_{2,26}$--26$_{2,25}$ A, $v_t = 1$ & 299.131294 & $-3.397$ & 110 & 397.2 \\
24$_{8,16}$--23$_{8,16}$ E, $v_t = 0$ & 299.131361 & $-4.932$ & 98 & 220.4 \\
27$_{1,26}$--26$_{1,25}$ A, $v_t = 1$ & 299.133165 & $-3.397$ & 110 & 397.2 \\
27$_{2,26}$--26$_{1,25}$ A, $v_t = 1$ & 299.135510 & $-4.243$ & 110 & 397.2 \\
12$_{6,6}$--11$_{5,7}$ A, $v_t = 0$ & 299.162434 & $-4.481$ & 50 & 70.0 \\
24$_{7,18}$--23$_{7,17}$ E, $v_t = 0$ & 300.020579 & $-3.426$ & 98 & 210.9 \\
24$_{7,18}$--23$_{7,17}$ A, $v_t = 0$ & 300.035621 & $-3.425$ & 98 & 210.9 \\
24$_{6,19}$--23$_{6,18}$ E, $v_t = 0$ & 300.046229 & $-3.415$ & 98 & 202.7 \\
24$_{6,19}$--23$_{6,18}$ A, $v_t = 0$ & 300.063893 & $-3.415$ & 98 & 202.7 \\
26$_{3,24}$--25$_{3,23}$ E, $v_t = 0$ & 300.070744 & $-3.398$ & 106 & 206.8 \\
34$_{2,32}$--34$_{1,33}$ A, $v_t = 0$ & 300.071349 & $-4.903$ & 138 & 340.3 \\
34$_{3,32}$--34$_{2,33}$ A, $v_t = 0$ & 300.072325 & $-4.903$ & 138 & 340.3 \\
26$_{3,24}$--25$_{3,23}$ A, $v_t = 0$ & 300.079079 & $-3.398$ & 106 & 206.8 \\
\hline
\multicolumn{5}{c}{Methanol (\methanol)} \\
\hline
4$_{-3,2}$--5$_{-2,4}$ E, $v_t = 0$ & 288.705624 & $-4.734$ & 36 & 70.9 \\
6$_{0,6}$--5$_{0,5}$ E, $v_t = 0$ & 289.939377 & $-3.976$ & 52 & 61.8 \\
6$_{1,6}$--5$_{1,5}$ E, $v_t = 0$ & 290.069747 & $-3.987$ & 52 & 54.3 \\
6$_{0,6}$--5$_{0,5}$ A, $v_t = 0$ & 290.110637 & $-3.975$ & 52 & 48.7 \\
6$_{5,2}$--5$_{5,1}$ E, $v_t = 0$ & 290.117786 & $-4.489$ & 52 & 172.8 \\
6$_{-5,1}$--5$_{-5,0}$ E, $v_t = 0$ & 290.138889 & $-4.489$ & 52 & 184.8 \\
6$_{5,2}$--5$_{5,0}$ A, $v_t = 0$ & 290.145085 & $-4.487$ & 52 & 186.6 \\
6$_{5,1}$--5$_{5,1}$ A, $v_t = 0$ & 290.145085 & $-4.487$ & 52 & 186.6 \\
6$_{4,3}$--5$_{4,2}$ A, $v_t = 0$ & 290.161348 & $-4.229$ & 52 & 129.1 \\
6$_{4,2}$--5$_{4,1}$ A, $v_t = 0$ & 290.161352 & $-4.229$ & 52 & 129.1 \\
6$_{4,3}$--5$_{4,2}$ E, $v_t = 0$ & 290.162356 & $-4.230$ & 52 & 136.6 \\
6$_{-4,2}$--5$_{-4,1}$ E, $v_t = 0$ & 290.183289 & $-4.227$ & 52 & 144.7 \\
6$_{2,5}$--5$_{2,4}$ A, $v_t = 0$ & 290.184674 & $-4.023$ & 52 & 86.5 \\
6$_{3,4}$--5$_{3,3}$ A, $v_t = 0$ & 290.189515 & $-4.100$ & 52 & 98.5 \\
6$_{3,3}$--5$_{3,2}$ A, $v_t = 0$ & 290.190549 & $-4.100$ & 52 & 98.5 \\
6$_{3,3}$--5$_{3,2}$ E, $v_t = 0$ & 290.209695 & $-4.097$ & 52 & 111.5 \\
6$_{-3,4}$--5$_{-3,3}$ E, $v_t = 0$ & 290.213180 & $-4.098$ & 52 & 96.5 \\
6$_{-1,5}$--5$_{-1,4}$ E, $v_t = 0$ & 290.248685 & $-3.975$ & 52 & 69.8 \\
6$_{2,4}$--5$_{2,3}$ A, $v_t = 0$ & 290.264068 & $-4.022$ & 52 & 86.5 \\
6$_{2,4}$--5$_{2,3}$ E, $v_t = 0$ & 290.307281 & $-4.024$ & 52 & 74.7 \\
6$_{-2,5}$--5$_{-2,4}$ E, $v_t = 0$ & 290.307738 & $-4.029$ & 52 & 71.0 \\
\hline
\multicolumn{5}{c}{Acetaldehyde (\acetaldehyde)} \\
\hline
16$_{0,16}$--15$_{0,15}$ E, $v_t = 0$ & 299.174780 & $-3.038$ & 66 & 123.6 \\
16$_{0,16}$--15$_{0,15}$ A, $v_t = 0$ & 299.218175 & $-3.038$ & 66 & 123.5 \\
15$_{6,10}$--14$_{6,9}$ A, $v_t = 0$ & 289.149836 & $-3.157$ & 62 & 192.2 \\
15$_{6,9}$--14$_{6,8}$ A, $v_t = 0$ & 289.149853 & $-3.157$ & 62 & 192.2 \\
15$_{6,9}$--14$_{6,8}$ E, $v_t = 0$ & 289.163254 & $-3.157$ & 62 & 192.2 \\
15$_{6,10}$--14$_{6,9}$ E, $v_t = 0$ & 289.200782 & $-3.157$ & 62 & 192.1 \\
7$_{2,5}$--6$_{1,6}$ E, $v_t = 0$ & 289.238518 & $-4.459$ & 30 & 35.1 \\
15$_{5,11}$--14$_{5,10}$ A, $v_t = 0$ & 289.272555 & $-3.132$ & 62 & 167.5 \\
15$_{5,10}$--14$_{5,9}$ A, $v_t = 0$ & 289.273794 & $-3.132$ & 62 & 167.5 \\
15$_{5,10}$--14$_{5,9}$ E, $v_t = 0$ & 289.316721 & $-3.132$ & 62 & 167.4 \\
15$_{5,11}$--14$_{5,10}$ E, $v_t = 0$ & 289.322537 & $-3.132$ & 62 & 167.4 \\
15$_{4,12}$--14$_{4,11}$ A, $v_t = 0$ & 289.506111 & $-3.112$ & 62 & 147.2 \\
15$_{4,11}$--14$_{4,10}$ A, $v_t = 0$ & 289.558746 & $-3.112$ & 62 & 147.3 \\
15$_{4,11}$--14$_{4,10}$ E, $v_t = 0$ & 289.562297 & $-3.112$ & 62 & 147.2 \\
15$_{4,12}$--14$_{4,11}$ E, $v_t = 0$ & 289.585326 & $-3.112$ & 62 & 147.2 \\
15$_{3,13}$--14$_{3,12}$ A, $v_t = 0$ & 289.602928 & $-3.097$ & 62 & 131.5 \\
15$_{3,13}$--14$_{3,12}$ E, $v_t = 0$ & 289.716267 & $-3.101$ & 62 & 131.4 \\
7$_{2,5}$--6$_{1,6}$ A, $v_t = 0$ & 289.942040 & $-4.371$ & 30 & 35.0
\enddata
\tablecomments{For (tentatively) detected species (\dimethylether and \methylformate), all transitions which could be responsible for the observed signals are listed. For non-detected species (\methanol and \acetaldehyde), only strong transitions ($\log_{10}A_\mathrm{ul}\,(\mathrm{s}^{-1})\ge-5$ and $E_\mathrm{u} \le 200$\,K) that effectively constrain the colum upper limits are listed.}
\end{deluxetable*}

\section{Parametric Disk Model}\label{appendix:toy_model}
% The high excitation temperature and broad line width derived in Section \ref{subsec:spectral_analysis} suggest that the emission traces the warm component in the innermost region of the disk. However, the exact radial distributions of the emitting molecules, which are spatially unresolved, are unclear. In addition, the detection of \dimethylether while the non-detection of \methanol, which result in an unusually high column density ratio of \dimethylether/\methanol, may indicate that the spatial distributions of \methanol are more compact than that of \dimethylether. 
Here, we demonstrate a detailed radiative transfer calculation of a disk model that aims to constrain what spatial distributions of COMs in the inner disk can explain the observed \dimethylether emission with a more realistic disk physical structure. Additionally, we examine if the difference in the spatial distributions of \dimethylether and \methanol can explain the non-detection of \methanol, as discussed in Section \ref{subsec:chemical_origin_of_COMs}.   

\begin{figure*}
\epsscale{1.15}
\plotone{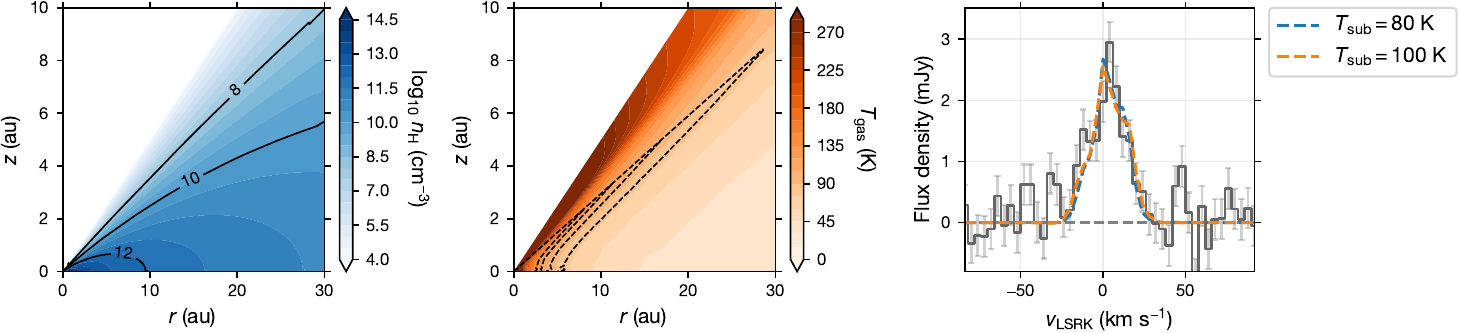}
\caption{Left and middle: Gas density (left) and temperature (middle) structure of the disk model. The dashed lines in the middle panel enclose the region where molecules distribute for each boundary temperature in the models (80, 100, 120, and 140\,K from outside). The upper boundary of the regions is $z/r = 0.3$, which is also depicted by the dashed lines. Right: Comparison of the observed and modeled spectra of the \dimethylether transitions at $\sim299.89$\,GHz for the models with different boundary temperatures of \dimethylether (blue for 80\,K model and orange for 100\,K model). The spectra of these models are nearly identical and overlap.}
\label{fig:model_comparison}
\end{figure*}

As the disk density and temperature structures, we used the disk model tailored to MWC~480 in \citet{Zhang2021}. The details of the model are referred to the original paper. Briefly, the model assumed two populations (small and large) of dust grains, and the distributions of large dust grains are constrained from the high-resolution ALMA observations. The small dust grains are assumed to be completely coupled with the gas. While in the original model the gas temperature structure is computed by the thermochemical modeling, given the uncertainty of the modeling in the inner region, here we simply assumed that the gas temperature is the same as the dust temperature (surface-area-weighted over both populations) computed with thermal Monte Carlo simulations using the RADMC-3D package \citep{RADMC3D}. The gas density and temperature distributions used for the present calculations are shown in Figure \ref{fig:model_comparison}.

The distributions of the warm \dimethylether and \methanol gas are assumed based on the dust temperature, which controls the ice sublimation. Since in general the sublimation temperature of a specific molecule is proportional to the binding energy of the molecule \citep[e.g.,][]{Furuya2014}, the sublimation temperature ratio between \dimethylether and \methanol should be the same as the binding energy ratio. Although the absolute values of binding energy for \dimethylether and \methanol shows some scatter depending on the experimental methods and the composition of the surface, there are a trend that \dimethylether have a lower binding energy than \methanol. The experimental estimates of the binding energies of \dimethylether (${\approx}4000$\,K; \citealt{Lattelais2011}) and \methanol (${\approx}5000$--5500\,K; \citealt{Ferrero2020, Minissale2022}) thus lead to the sublimation temperature ratio of $T_\mathrm{sub}(\mathrm{CH_3OH})/T_\mathrm{sub}(\mathrm{CH_3OCH_3})\approx1.2$--$1.4$. Based on this ratio and the gas density of the inner disk, we constructed a series of models, in which \dimethylether and \methanol distribute with a constant abundance $x$ (relative to H$_2$ molecule) in the regions where the dust temperature is higher than 80 and 100\,K for \dimethylether, and 100, 120, and 140\,K for \methanol. This emulates the drastic gas-phase abundance increases due to the thermal desorption. We additionally set an upper boundary of $z/r\lesssim0.3$, reflecting the expectation that molecules can be destroyed in the irradiated surface of the disk \citep[e.g.,][]{Walsh2014}. Outside these boundaries, the abundances of \methanol and \dimethylether are set to zero, which is a suitable first-order approximation given the steep (exponential) dependence of gaseous molecular abundances on visual extinction and dust temperature around its sublimation temperature. The dependence of COM abundances on other parameters such as gas density is weaker \citep[e.g.,][]{Hasegawa1992, Furuya2014}. This resulted in the distributions of warm \dimethylether and \methanol gas all within ${\lesssim}20$--30\,au (the middle panel of Figure \ref{fig:model_comparison}). The velocity structure of the warm gas is assumed to be in Keplerian rotation with a known stellar mass of $2.1M_\odot$ \citep[e.g.,][]{Teague2021_MAPS}. 

For the models with different combinations of boundary temperatures as listed in Table \ref{tab:toy_models}, which have at maximum a sublimation temperature ratio of 1.5 as the most extreme case, we compute the image cubes covering the frequency ranges of the robustly detected \dimethylether transitions at $\sim$\,299.89\,GHz and the undetected \methanol transitions at 290.0--290.4\,GHz using the RADMC-3D package \citep{RADMC3D}. The resulting image cubes are convolved with the observing beam, continuum-subtracted, and spatially integrated to create the disk-integrated spectra, which is then compared to the observed spectra. The abundance $x$(\dimethylether) and $x$(\methanol) are adjusted to fit the observed spectra.
% Since the observed emission is compact and spatially unresolved, the spatial filtering effect in the observations is expected to be negligible. During the modeling, we found that the other tentatively detected transitions and undetected transitions covered in other frequency ranges are not relevant to constrain the molecular distributions, and thus focused on the aforementioned \dimethylether transitions and undetected \methanol transitions.

The right panel of Figure \ref{fig:model_comparison} compares the best-fit model with the observed spectrum of \dimethylether at ${\sim}299.89$\,GHz. Both models with boundary temperatures of 80 and 100\,K reproduce the spectrum well, particularly the broad line width. Furthermore, the non-detection of \methanol places constraints on the abundance ratio $x$(\dimethylether)/$x$(\methanol) of the warm gas, as listed in Table \ref{tab:toy_models}. The constrained lower limits of the abundance ratios span ${\sim}0.3$--2,  depending on the combination of the boundary temperatures. These values are lower than the column density ratio derived in the spectral analysis in Section \ref{subsec:spectral_analysis} (where the same distributions are implicitly assumed) by a factor of ${\sim}3$--20. This indicates that the spatial distribution is a critical factor in measuring the correct abundance ratio. Further spatially-resolved observations are necessary. It should be noted that the disk physical structure used in this modeling is an extrapolation of the outer disk structure. The actual structure can vary due, for example, to the accretion heating that could modify the temperature structure of the inner disk. Constraining the inner disk structure is also essential to infer the inner disk chemistry.

\begin{deluxetable*}{ccccccccc}
\label{tab:toy_models}
\tablecaption{Disk Modeling Results}
\tablehead{\colhead{$r_\mathrm{in}$} & \colhead{$r_\mathrm{out}$} & \colhead{$(z/r)_\mathrm{min}$} & \colhead{$(z/r)_\mathrm{max}$} & \colhead{$T_\mathrm{sub}(\mathrm{CH_3OCH_3})$} & \colhead{$T_\mathrm{sub}(\mathrm{CH_3OH})$} & \colhead{$x(\mathrm{CH_3OCH_3})$} & \colhead{$x(\mathrm{CH_3OH})$} & \colhead{$x(\mathrm{CH_3OCH_3})/x(\mathrm{CH_3OH})$} \\ \colhead{(au)} & \colhead{(au)} & \colhead{} & \colhead{} & \colhead{(K)} & \colhead{(K)} & \colhead{} & \colhead{} & \colhead{}}
\startdata
% R1 & 10  & 20 & 0.2 & 0.3 & --- & --- & $2.3\times10^{-8}$ & --- & --- \\
% R2 & 0.5 & 10 & 0.2 & 0.3 & --- & --- & $1.9\times10^{-7}$ & --- & --- \\
0.5 & 30 & 0.0 & 0.3 & 80  & 100 & $3.0\times10^{-8}$ & $<3.1\times10^{-8}$ & $>0.97$\\
0.5 & 30 & 0.0 & 0.3 & 80  & 120 & $3.0\times10^{-8}$ & $<1.2\times10^{-7}$ & $>0.25$\\
% T1c & 0.5 & 30 & 0.0 & 0.3 & 80  & 140 & $3.0\times10^{-8}$ & $<4.4\times10^{-7}$ & $>0.068$\\
0.5 & 30 & 0.0 & 0.3 & 100 & 120 & $1.8\times10^{-7}$ & $<1.0\times10^{-7}$ & $>1.8$ \\
0.5 & 30 & 0.0 & 0.3 & 100 & 140 & $1.8\times10^{-7}$ & $<5.8\times10^{-7}$ & $>0.31$ \\
\enddata
\end{deluxetable*}

\section{Detection of a new submillimeter source}\label{appendix:new_submm_detection}
\begin{figure}
% \epsscale{1.2}
\plotone{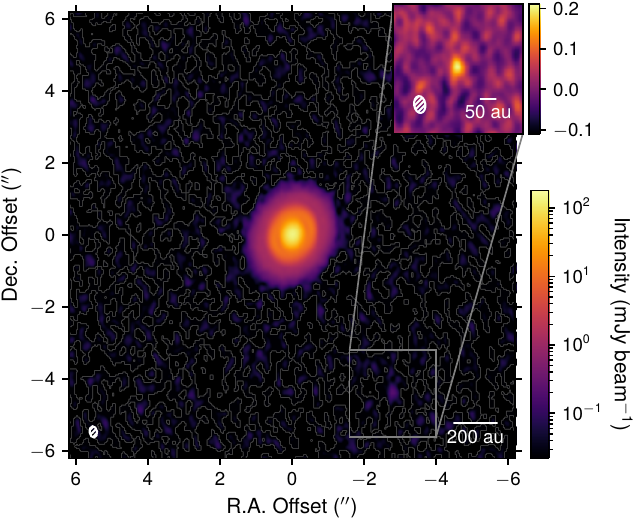}
\caption{1.03\,mm continuum image made with a Briggs robust value of 0.5. The inset panel presents the zoomed-in view of the vicinity of the new submillimeter source. The beam size of $0\farcs31\times0\farcs21\,(\mathrm{P.A.}=8\fdg2)$ is shown in the lower left corner of each panel. A scale bar is shown in the lower right corner of each panel.}
\label{fig:continuum}
\end{figure}

We report the detection of a new submillimeter source at a separation of ${\approx}5\farcs2$ from the central star with a position angle of $212\fdg5$ (Figure \ref{fig:continuum}). The emission is compact and reaches a peak S/N of ${\approx}7$ on the image made with a Briggs robust value of 0.5. The coordinate of the continuum peak is $\alpha(\mathrm{J2000}) = 4^\mathrm{h}58^\mathrm{m}46\fs066$; $\delta(\mathrm{J2000}) = 29^\mathrm{d}50^\mathrm{m}32\fs04$. To characterize the emission, we performed a Gaussian fit to this component with the CASA task \texttt{imfit}, which obtained a deconvolved size of $0\farcs33\times0\farcs13\,(\mathrm{P.A.}=179\arcdeg)$ and a flux density of 0.32 mJy. Although the origin of this emission is unclear, this component is located at outside the $^{12}$CO gas disk analysed in \citet{Teague2021_MAPS}, and thus it is unlikely that the emission is attributed to the circumplanetary disk around a forming planet in the MWC~480 disk. It is possible that this originates from the disk or envelope around a field star, or a background galaxy.

\bibliography{sample631}{}
\bibliographystyle{aasjournal}



\end{document}